\documentclass[12pt,draftclsnofoot,onecolumn]{IEEEtran}
\usepackage{amsmath,amssymb,amsfonts}
\usepackage{algorithmic}
\usepackage{algorithm}
\usepackage{array}
\usepackage[caption=false,font=footnotesize]{subfig}
\usepackage{textcomp}
\usepackage{stfloats}
\usepackage{url}
\usepackage{verbatim}
\usepackage{graphicx}
\usepackage{wrapfig}
\usepackage{cite}
\usepackage{booktabs}

\usepackage[colorlinks=true,
            linkcolor=blue,   
            citecolor=blue,  
            urlcolor=black]{hyperref}
\hypersetup{pdfborder={0 0 0}}
\usepackage[normalem]{ulem} %
\usepackage{soul} %

\usepackage{xcolor} %
\usepackage{orcidlink}          
\usepackage{tikz}
\usetikzlibrary{arrows.meta,positioning,calc}
\newif\ifincludeauthorbiographies
\includeauthorbiographiestrue
\begin{document}



\title{Enhanced ShockBurst for Ultra-Low-Power On-Demand Sensing}


\author{Ziyao Zhou\orcidlink{0009-0007-0671-9947},
        Chen Shen\orcidlink{0009-0008-4221-2534},
        Tiancheng Cao\orcidlink{0000-0002-7259-5192},
        and Hen-Wei Huang\orcidlink{0000-0003-1921-8897}%
\thanks{This work was supported by the Nanyang Assistant Professorship, the MOE Tier 1 grants RG71/24 and RG26/25, and the MTC MedTech Programmatic Fund M24N9b0125. \textit{(Corresponding author: Chen Shen.)}}%
\thanks{Ziyao Zhou, Chen Shen, Tiancheng Cao, and Hen-Wei Huang are with the School of Electrical and Electronic Engineering, Nanyang Technological University, Singapore (e-mail: ZHOU0557@e.ntu.edu.sg; shenchen@ntu.edu.sg; tiancheng.cao@ntu.edu.sg; henwei.huang@ntu.edu.sg). Tiancheng Cao is also with Harvard Medical
School, Boston, MA, USA, and the Department of Mechanical Engineering, Massachusetts Institute of Technology, Cambridge, MA 02139, USA. Hen-Wei Huang is also with the Lee Kong Chian School of Medicine, Nanyang Technological University, Singapore.}%
}


\maketitle

\begin{abstract}
On-demand sensing requires battery-powered Internet-of-Things (IoT) and implantable medical devices to remain in deep sleep and activate wireless communication only when data transmission is required. In such systems, battery lifetime depends strongly on radio active time. This work investigates how communication architecture and physical layer (PHY) configuration influence radio active time by comparing connection-oriented Bluetooth Low Energy (BLE) with connectionless Enhanced ShockBurst (ESB) on identical BLE-compatible hardware. Under identical 2 Mbps PHY configurations, ESB reduces wake-up latency and energy consumption to approximately one-twentieth of BLE by eliminating connection establishment and maintenance overhead. Increasing the ESB PHY rate from 2 to 4 Mbps further shortens packet airtime by approximately 52\% and reduces transmission energy by approximately 43\%. Finally, a first-in, first-out (FIFO)-triggered implantable loop recorder prototype demonstrates that jointly optimizing communication architecture, PHY configuration, and buffered transmission enables sleep--wake operation and reduces total system power consumption by approximately 60\% compared with conventional BLE operation. These results identify minimizing radio active time as a key design principle for ultra-low-power on-demand sensing and provide practical guidance for battery-powered sensing systems.

\end{abstract}


\section{Introduction}
Wireless sensor networks (WSN) have become a fundamental component of modern Internet-of-Things (IoT) systems, enabling intelligent sensing across healthcare, industrial automation, environmental monitoring, and smart infrastructure~\cite{Koulouras2025Evolution, Tosi2017Performance, Jamshed2022Challenges}. As sensing devices become increasingly miniaturized and battery-powered, extending battery lifetime has become one of the primary design challenges\cite{Anastasi2009Energy}. Considerable efforts have therefore focused on reducing unnecessary sensing, computation, and wireless communication. Among these approaches, on-demand sensing has emerged as an attractive operating paradigm, where sensing devices remain in an ultra-low-power sleep state and activate sensing and wireless communication only when data transmission is explicitly requested or triggered by an event \cite{9900377, 9204414}. 

Unlike continuous wireless monitoring, on-demand sensing imposes a fundamentally different set of communication requirements. During extended idle periods, the wireless node should consume virtually no power while remaining capable of responding immediately when sensing and communication are required. Therefore, an ideal protocol should support rapid wake-up, minimal communication overhead, reliable packet delivery, and immediate return to the lowest-power operating state after transmission~\cite{sutton2019blitz}. Consequently, minimizing the duration for which the radio remains active—including connection establishment, packet transmission, and unnecessary communication events—becomes one of the most important design objectives for ultra-low-power sensing systems. These requirements are particularly important in medical devices~\cite{Darwish2011Wearable, Abdigazy2024EndToEndIE, Traverso2026MiniaturizedIE}, such as implantable loop recorders (ILRs)~\cite{bisignani2019implantable}, ingestible sensors, and other long-term physiological monitoring systems, where communication events occur infrequently but timely access to stored physiological data is critical. 

Bluetooth Low Energy (BLE) has become the de facto wireless communication standard for low-power IoT and medical devices because of its widespread interoperability and standardized ecosystem. Although BLE provides an advertising mode that avoids connection establishment, BLE advertising does not support bidirectional data exchange, and its packets remain prone to loss because it lacks acknowledgment (ACK)-based automatic retransmission~\cite{10106496}. This limitation makes advertising unsuitable for applications requiring reliable delivery of critical data. Reliable communication therefore generally relies on BLE's connection-oriented mode, in which devices periodically exchange packets to establish and maintain a communication link. Although this architecture provides reliable bidirectional communication, it inevitably introduces connection establishment, service discovery, scheduled connection events, and continuous link-maintenance overhead~\cite{cho2014analysis}. Such mechanisms are highly beneficial for continuous data streaming but become increasingly inefficient for event-driven sensing applications, where communication occurs only sporadically. Consequently, BLE may spend substantially more energy maintaining communication readiness than transmitting the sensing data itself. Furthermore, BLE supports a maximum physical layer (PHY) rate of 2 Mbps~\cite{Badihi2020On}, which may limit opportunities to reduce packet airtime through higher-rate transmission. Despite the widespread adoption of BLE, the extent to which its connection-oriented architecture limits on-demand sensing has not been systematically investigated.

Connectionless communication offers an alternative architectural paradigm. Instead of maintaining an active communication session, devices remain completely inactive until communication is required, thereby eliminating continuous connection-maintenance overhead. Nordic Semiconductor's Enhanced ShockBurst (ESB) is one representative implementation of this architecture~\cite{veilleux2018esb}. Unlike BLE, ESB enables direct packet transmission immediately after wake-up while supporting configurable acknowledgment and retransmission mechanisms for reliable communication~\cite{11337444}. Importantly, ESB is fully compatible with the same Nordic wireless hardware platform as BLE, allowing firmware-level switching between the two protocols without radio-frequency (RF) hardware modification. This hardware compatibility not only enables a fair protocol-level comparison by eliminating hardware-induced bias but also provides a practical migration path for existing BLE-based sensing systems. Beyond communication architecture, radio active time can be further reduced by increasing the PHY rate to 4 Mbps to shorten packet airtime and by aggregating sensor data into burst transmissions, thereby reducing the frequency of radio activation. Together, these techniques provide promising approaches for minimizing wireless communication energy in on-demand sensing systems.

\begin{figure*}[!htbp]
    \centering
    \includegraphics[width=0.78\textwidth]{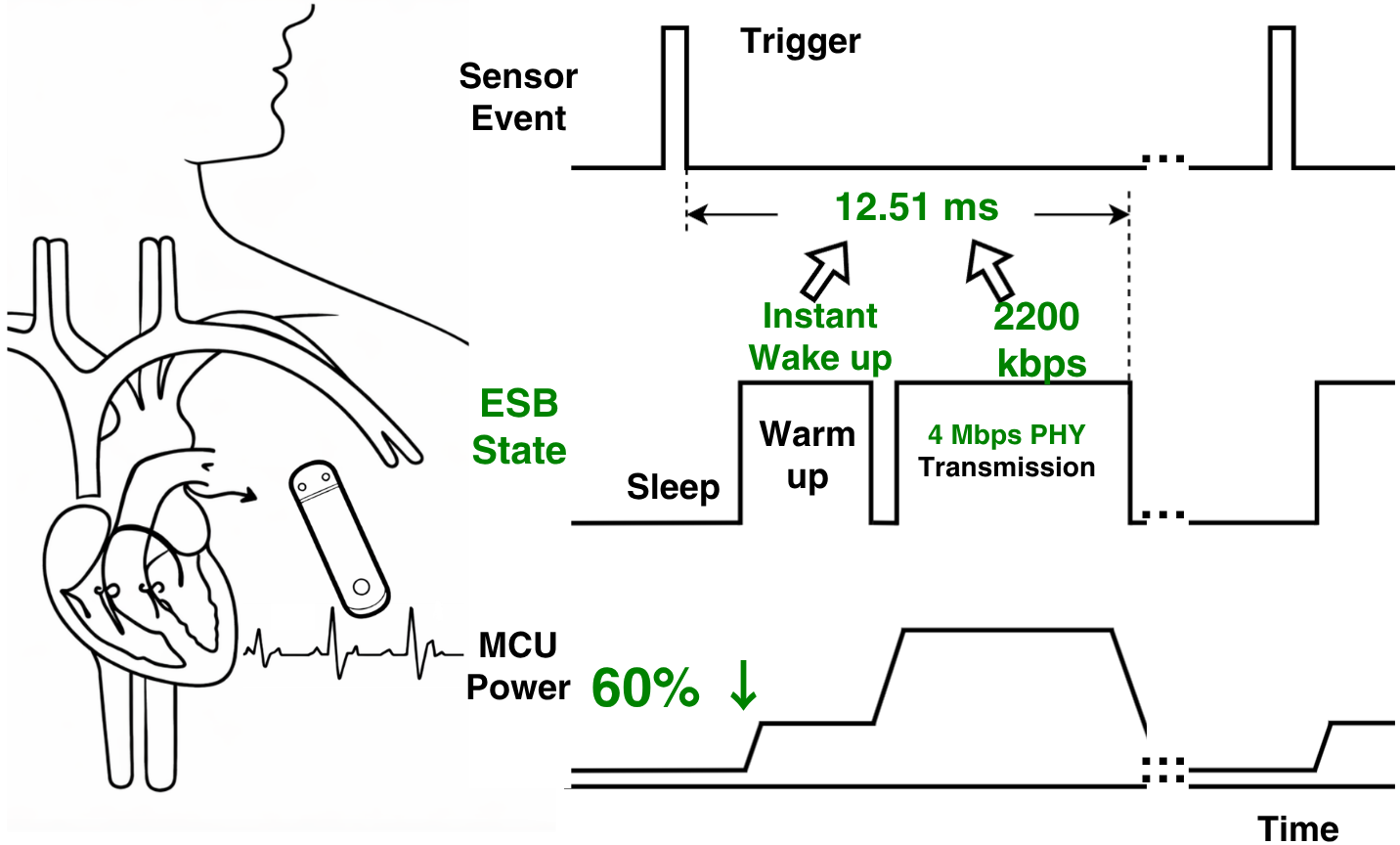}
    \caption{Conceptual overview of the FIFO-triggered ILR prototype and its ESB sleep--wake operation. ECG samples are buffered while the MCU remains in the system-off state; a FIFO-threshold event initiates wake-up and transmission over the ESB 4 Mbps PHY. Under the evaluated configuration, the prototype achieved a 12.51 ms wake-up-to-first-packet latency and approximately 60\% lower total system power consumption than the BLE continuous-connection baseline.}
    \label{fig:prototype_overview}
\end{figure*}

Motivated by these architectural differences, we systematically investigate the suitability of connection-oriented and connectionless wireless communication architectures for ultra-low-power on-demand sensing. Rather than proposing a new wireless protocol, we use BLE and ESB as representative implementations to elucidate how communication architecture influences wake-up latency, energy consumption, throughput, reliability, and practical sensing performance. To isolate protocol-level differences, both protocols are implemented on the same wireless hardware platform and initially configured with the identical 2 Mbps PHY. Building upon this comparison, we investigate how increasing the ESB PHY rate shortens packet airtime and reduces transmission energy, and finally evaluate the resulting design in the ILR prototype shown in Fig.~\ref{fig:prototype_overview}. In this prototype, the MAX30003 buffers electrocardiogram (ECG) samples while the MCU remains in the system-off state. Reaching the FIFO threshold wakes the MCU, initiates SPI readout, and triggers burst transmission over the ESB 4 Mbps PHY, thereby emulating on-demand waveform retrieval. Collectively, the proposed communication architecture, PHY selection, and buffering strategy optimize radio active time, enabling substantial system-level energy savings for battery-powered sensing devices.

The main contributions of this work are summarized as follows:

\begin{itemize}

\item We establish the impact of wireless communication architecture on on-demand sensing, demonstrating that connection-oriented communication introduces substantial latency and energy overhead due to connection establishment and maintenance, whereas connectionless communication enables true sleep–wake operation.

\item We systematically investigate multiple approaches for reducing radio active time, including connectionless communication, higher PHY rates, and FIFO-based burst communication, and quantify their respective impacts on communication latency, transmission energy, throughput, and overall system power consumption.

\item We experimentally characterize the design trade-offs of connectionless communication, including retransmission configuration, reverse-link communication, and PHY selection, demonstrating that higher transmission rates reduce radio airtime and transmission energy in addition to improving communication throughput.

\item We validate the proposed design principles using a FIFO-triggered ILR, demonstrating that co-optimizing communication architecture, transmission airtime, and radio activation mechanism reduces total system power consumption by approximately 60\% compared with conventional BLE continuous-connection operation.

\end{itemize}

\section{Protocol Comparison Methodology}

Both BLE and ESB can run on the same Nordic wireless microcontroller unit (MCU) with identical hardware. To enable a fair and interpretable comparison, their differences should therefore be first examined at the software level for a single transmission. Specifically, each protocol adopts a distinct procedure for data processing, scheduling, and transmission, which directly determines the measured performance metrics.

\begin{figure}[!htbp]
    \centering
    \includegraphics[width=0.72\columnwidth]{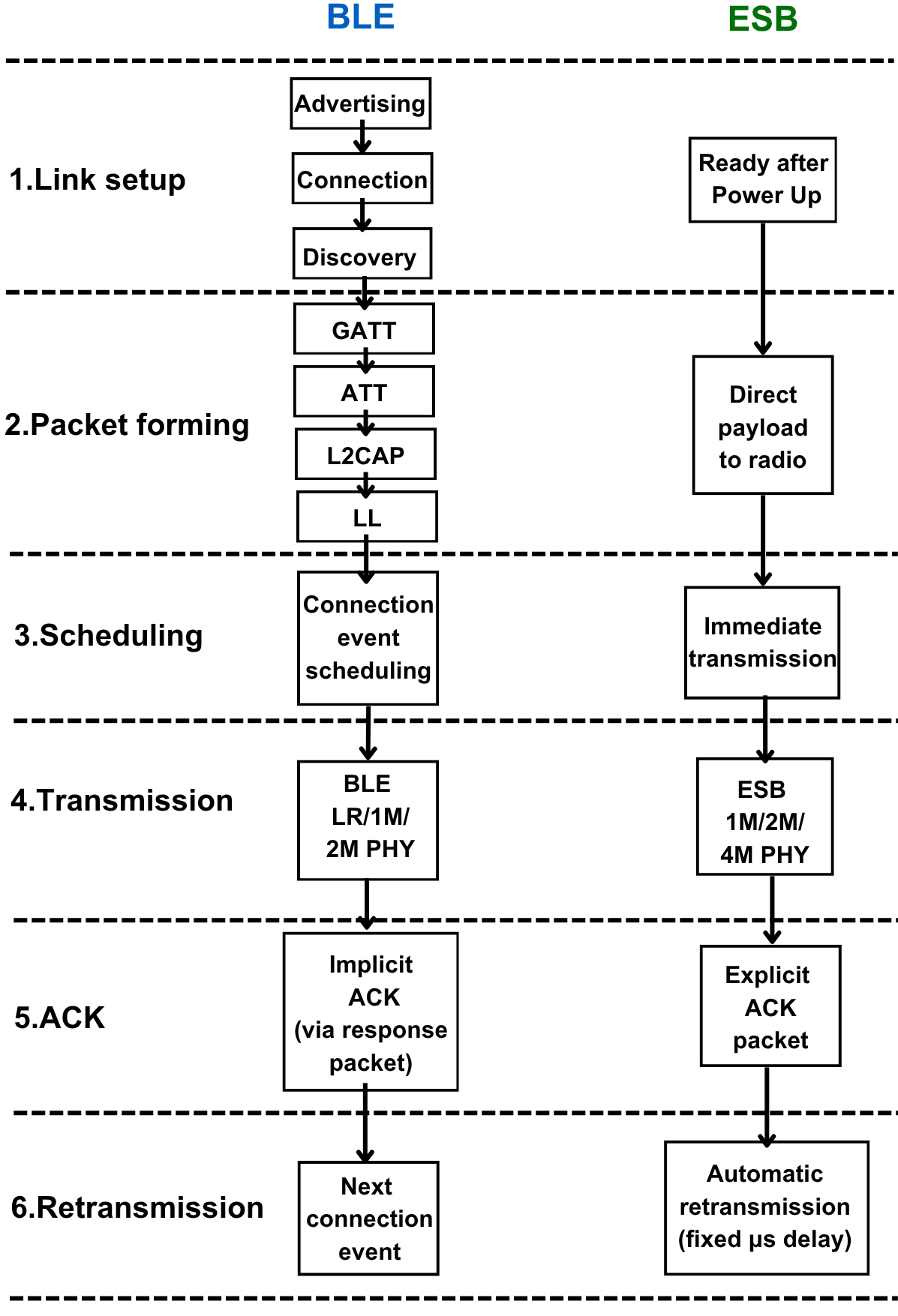}
    \caption{Comparison of single transmission procedures between BLE and ESB. BLE relies on connection-oriented scheduling, whereas ESB enables connectionless execution with lightweight packet handling.}
    \label{fig:ble_esb_flow}
\end{figure}

As illustrated in Fig.~\ref{fig:ble_esb_flow}, BLE and ESB follow different procedures for a single transmission. BLE requires a link establishment process, including advertising, connection, and service discovery, before data transmission, whereas ESB can directly enter the transmission state.

During packet formation, BLE processes data through the Generic Attribute Profile (GATT), Attribute Protocol (ATT), Logical Link Control and Adaptation Protocol (L2CAP), and link layer, while ESB directly forwards the payload to the radio. For scheduling, BLE transmissions are constrained by connection events, whereas ESB supports immediate transmission without predefined time slots.

BLE supports a maximum PHY rate of 2 Mbps, whereas ESB supports up to 4 Mbps and can therefore achieve a higher transmission speed. However, to ensure a fair protocol-level comparison, both BLE and ESB were configured to operate with the 2 Mbps PHY in the first experiments. In terms of acknowledgment, BLE employs an implicit Link Layer acknowledgment mechanism within connection events, whereas ESB uses explicit ACK packets. For retransmission, BLE retransmits an unacknowledged packet at a subsequent transmission opportunity scheduled by its connection-event procedure~\cite{11222604}. In contrast, ESB performs automatic retransmission after a configurable microsecond-level delay, which was set to the implementation default of 600~$\mu$s in this study.

\subsection{Sleep--Wake Evaluation}

Long-term IoT monitoring systems often adopt an on-demand sensing strategy, in which the device remains in an ultra-low-power sleep state for extended periods and wakes only briefly to acquire and transmit data. This duty-cycled operation effectively reduces the average power consumption and prolongs battery lifetime. However, BLE is less suited to such an operation because each wake-up event requires a relatively long warm-up period to re-establish connection, introducing additional delay and energy overhead. In contrast, ESB operates in a connectionless manner and therefore enables significantly faster wake-up, making it more suitable for on-demand IoT applications.

To evaluate BLE and ESB under on-demand monitoring, both transmitters were configured to wake from sleep every 10 s, transmit one packet, and then return to the sleep state. For a fair comparison, both protocols used a 244-byte payload, corresponding to the maximum application payload supported by the BLE configuration while remaining below the 252-byte ESB payload limit.

Throughout the experiments, both the transmitter and receiver were placed inside an RF-shielded enclosure. The transmitter was powered and monitored using a Power Profiler Kit II (PPK2), which sampled the supply current at 100 kSa/s to capture the instantaneous power profile~\cite{nordicsemi_ppk2_userguide_v1.0.1}. For each BLE and ESB wake-up cycle, the elapsed time and energy consumption from wake-up initiation through completion of the first-packet transmission were extracted from the PPK2 measurements. These measurements were then used to quantify the warm-up time and energy overhead of the two protocols under identical operating conditions.

\subsection{Bidirectional Evaluation}

Despite its potential advantage in wake-up latency, ESB lacks the symmetric bidirectional communication capability provided by BLE. BLE supports independent data transfer in both directions within each connection event~\cite{Gomez2012Overview}. In contrast, ESB adopts a master-initiated transaction model, where all communication is triggered by the transmitter. The receiver can only return data by embedding it within ACK packets, rendering the reverse link entirely dependent on forward-link activity.

To systematically characterize the similarities and differences between the two protocols in bidirectional communication, we adopted a methodology analogous to continuous packet transmission. Specifically, the packet generation rate of one node is controlled, while the peer node transmits in the reverse direction at its maximum achievable throughput under the current condition.

For BLE, the packet transmission rate at the peripheral is adjusted to regulate the throughput observed at the central device, with a fixed payload size of 244 bytes. Meanwhile, the central transmits packets of the same size back to the peripheral at its maximum achievable throughput under the given link condition. For ESB, the packet transmission rate of the transmitter is controlled to regulate the throughput at the receiver, using a payload size of 252 bytes. The receiver then responds by sending ACK packets at its maximum capability under the current configuration. To investigate the impact of reverse-link payload size, the ACK packet size was varied among 2 bytes, 132 bytes, and 252 bytes.


\subsection{Reliability Evaluation}

Communication reliability is crucial for IoT devices because sensing applications depend on the accurate and complete delivery of acquired data. A reliable communication protocol therefore requires ACK and retransmission mechanisms to detect unsuccessful transmissions and recover missing packets. These mechanisms enable the receiver to verify packet continuity and preserve the transmission order, thereby supporting accurate and effective IoT operation.

\begin{figure}[!htbp]
    \centering
    \includegraphics[width=0.72\columnwidth]{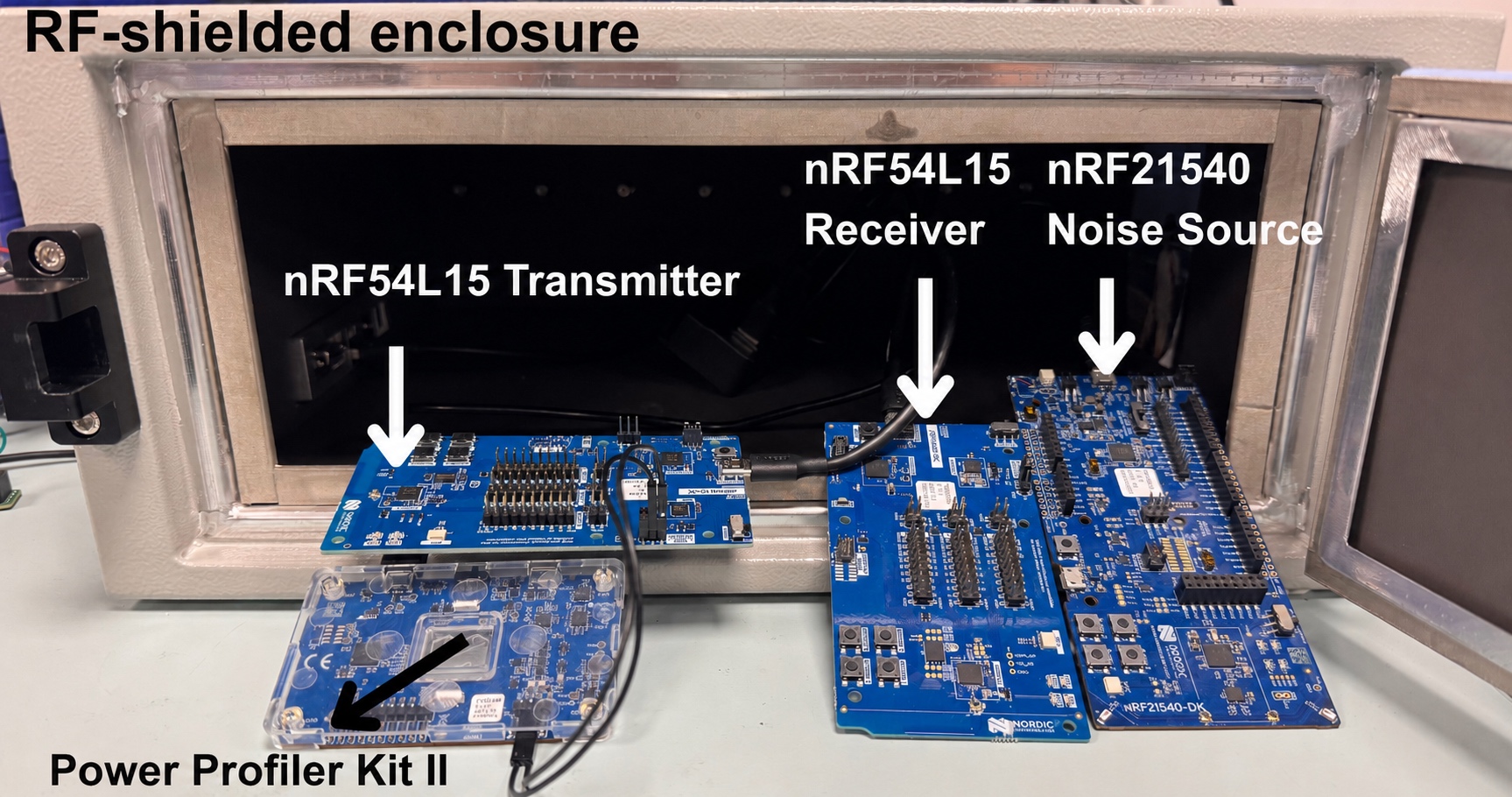}
    \caption{Experimental setup for the communication reliability evaluation inside the RF-shielded enclosure. The setup comprised an nRF54L15 transmitter and receiver, an nRF21540-based interference source, and a PPK2 for transmitter power measurement.}
    \label{fig:reliability_experimental_setup}
\end{figure}

To compare the communication reliability of ESB and BLE, each transmitter was configured to send sequentially numbered data packets. The receiver decoded the sequence number of each received packet and recorded the number of missing packets. As shown in Fig.~\ref{fig:reliability_experimental_setup}, an additional interference source was operated continuously at 20 dBm to generate random data and transmit at randomly selected frequencies within the 2.4 GHz industrial, scientific, and medical (ISM) band. The transmitter's transmission power (TXP) was manually varied across all supported settings from -28 to 8 dBm, and the packet loss rate was recorded at each setting.

ESB supports both ACK and automatic retransmission, and the maximum number of retransmission attempts can be configured. When the transmitter does not receive an ACK for a transmitted packet, it automatically retransmits that packet up to the configured limit. To characterize the effect of the retransmission count on communication reliability, the ESB retransmission count was configured as 1, 3, and 9. For each configuration and TXP setting, the packet loss rate, throughput, and system power consumption were recorded to quantify the trade-off between reliability and communication overhead.

BLE provides reliable data transfer intrinsically during a connection through its Link Layer acknowledgment and retransmission procedure. Each data-channel packet carries sequence information through the sequence number (SN) and next expected sequence number (NESN) fields. These fields enable the receiving device to acknowledge a successfully received packet, identify duplicate packets, and preserve packet order. If the acknowledgment state does not advance, the transmitting device automatically retransmits the same packet in a subsequent transmission opportunity until it is acknowledged or the connection is terminated~\cite{BluetoothCoreSpecV62}. Unlike ESB, however, BLE manages this retransmission procedure internally within the protocol stack and does not expose a fixed retransmission-count parameter to the application. In addition, the connection interval may affect communication reliability because it determines how frequently retransmission opportunities become available. Therefore, the BLE connection interval was varied among 7.5, 100, and 400 ms to evaluate how the available retransmission opportunities affect packet loss rate, throughput, and system power consumption under interference.

\section{PHY Comparison Methodology}

In addition to the 2 Mbps PHY used in the preceding protocol-level comparison, ESB provides a 4 Mbps PHY that supports a higher transmission rate. To compare the characteristics of the two PHYs, two sets of experiments were designed for single-packet and continuous-packet transmission, respectively. The single-packet experiment compared the power profiles, transmission times, and energy consumption associated with transmitting an identical packet. The continuous-packet experiments were conducted in the presence of the same controlled 2.4 GHz interference described above. They focused on packet loss rate, throughput, and system power consumption across the supported TXP settings, thereby characterizing the performance and reliability differences between the ESB 2 Mbps and 4 Mbps PHYs. This comparison was subsequently used to select the most suitable ESB PHY configuration for the on-demand sensing evaluation.

\section{System Evaluation Methodology}
To implement an ILR prototype for electrocardiogram (ECG) monitoring, the Analog Devices MAX30003 was selected as the ECG acquisition sensor and interfaced with the nRF54L15 wireless MCU to form a complete ILR prototype. During the experiments, the nRF54L15 and the MAX30003 were powered independently using two PPK2 units to enable precise current measurements. A global timestamp was enabled on both channels to facilitate post-processing data alignment. An additional nRF54L15 development kit (DK) was deployed as a receiver.

The sensor was configured to sample continuously at 128 samples per second to emulate a scenario in which abnormal ECG events trigger sustained data transmission. This sampling rate is sufficient to capture key ECG features and supports advanced analyses such as heart rate variability (HRV) assessment~\cite{Kwon2018Electrocardiogram}. To support the low-power requirement, the FIFO functionality of the MAX30003 was enabled. Fig.~\ref{fig:fifo_synchronized_on_demand}(a) illustrates the resulting control flow. Once a predefined number of words had accumulated in the FIFO, the MAX30003 generated an interrupt to awaken the MCU. The MCU then completed initialization, retrieved the buffered samples through a serial peripheral interface (SPI) batch read, transmitted the data using the ESB 4 Mbps PHY, and returned to the system-off state after receiving an ACK.

Fig.~\ref{fig:fifo_synchronized_on_demand}(b) shows the synchronized timing of FIFO accumulation and MCU activity. While the FIFO accumulated ECG samples, the MCU remained in the system-off state. Reaching the FIFO threshold simultaneously initiated the MCU active window, which comprised a short SPI-reading phase followed by RF transmission and ACK reception. After completing these operations, the MCU returned to the system-off state until the next FIFO interrupt. This synchronization minimized communication idle activity and thereby reduced the average system power consumption.

\begin{figure}[!htbp]
    \centering
    \resizebox{0.78\columnwidth}{!}{%
    \begin{tikzpicture}[
        >=Latex,
        flow/.style={draw=black!70, rounded corners=1.5pt, minimum width=2.05cm,
                     minimum height=0.68cm, align=center, font=\sffamily\scriptsize,
                     line width=0.65pt},
        sensor/.style={flow, fill=blue!9},
        trigger/.style={flow, fill=yellow!18},
        active/.style={flow, fill=green!12},
        radio/.style={flow, fill=orange!16},
        sleep/.style={flow, fill=black!6},
        link/.style={->, line width=0.8pt, draw=black!75},
        sync/.style={densely dashed, line width=0.65pt, draw=black!50}
    ]
    \node[font=\sffamily\small, anchor=west] at (-0.55,0.72) {(a)};
    \node[sensor]  (sample) at (0,0) {Continuous ECG\\sampling};
    \node[sensor]  (fifo)   at (2.45,0) {FIFO\\accumulation};
    \node[trigger] (thres)  at (4.90,0) {FIFO threshold\\reached};
    \node[trigger] (irq)    at (7.35,0) {GPIO interrupt\\generated};
    \node[active] (wake) at (7.35,-1.25) {MCU wake-up\\and initialization};
    \node[active] (spi)  at (4.90,-1.25) {SPI batch\\read};
    \node[radio]  (tx)   at (2.45,-1.25) {ESB 4 Mbps burst\\and ACK};
    \node[sleep]  (off)  at (0,-1.25) {Return to\\system off};
    \draw[link] (sample) -- (fifo);
    \draw[link] (fifo) -- (thres);
    \draw[link] (thres) -- (irq);
    \draw[link] (irq) -- (wake);
    \draw[link] (wake) -- (spi);
    \draw[link] (spi) -- (tx);
    \draw[link] (tx) -- (off);
    \draw[link, rounded corners=4pt] (off.west) -- ++(-0.35,0) |- (sample.west);

    \node[font=\sffamily\small, anchor=west] at (-0.55,-2.38) {(b)};
    \draw[draw=green!55!black, line width=1.35pt] (3.10,-2.36) -- (3.48,-2.36);
    \node[font=\sffamily\tiny, text=green!40!black, anchor=west]
          at (3.55,-2.36) {Wake + SPI read};
    \draw[draw=orange!85!black, line width=1.35pt] (5.38,-2.36) -- (5.76,-2.36);
    \node[font=\sffamily\tiny, text=orange!75!black, anchor=west]
          at (5.83,-2.36) {RF transmission + ACK};
    \node[font=\sffamily\scriptsize, anchor=east] at (-0.05,-3.02) {FIFO level};
    \node[font=\sffamily\scriptsize, anchor=east] at (-0.05,-4.18) {MCU state};
    \draw[->, line width=0.7pt] (0,-5.15) -- (8.55,-5.15)
          node[below, font=\sffamily\scriptsize] {Time};
    \draw[densely dashed, draw=orange!80!black, line width=0.65pt]
          (0,-2.68) -- (8.35,-2.68)
          node[right, font=\sffamily\tiny, text=orange!70!black] {Threshold};
    \draw[draw=blue!75!black, line width=1.3pt]
          (0.15,-3.38) -- (2.10,-2.68) -- (2.10,-3.38)
          -- (4.45,-2.68) -- (4.45,-3.38)
          -- (6.80,-2.68) -- (6.80,-3.38) -- (8.30,-2.90);
    \foreach \x in {2.10,4.45,6.80} {
        \draw[sync] (\x,-2.55) -- (\x,-4.92);
    }

    \draw[draw=black!35, line width=0.85pt] (0.15,-4.48) -- (8.30,-4.48);
    \foreach \a/\b/\c in {2.10/2.22/2.38,4.45/4.57/4.73,6.80/6.92/7.08} {
        \draw[draw=green!55!black, line width=1.35pt]
              (\a,-4.48) -- (\a,-3.88) -- (\b,-3.88);
        \draw[draw=orange!85!black, line width=1.35pt]
              (\b,-3.88) -- (\c,-3.88) -- (\c,-4.48);
    }

    \draw[<->, draw=black!45, line width=0.55pt] (2.40,-4.76) -- (4.43,-4.76)
          node[midway, below, font=\sffamily\tiny, text=black!60] {System off};
    \end{tikzpicture}}%
    \caption{FIFO-synchronized on-demand sensing and communication operation. (a) The FIFO threshold aligns MCU wake-up, batch data retrieval, ESB 4 Mbps transmission, and ACK reception. (b) Reaching the FIFO threshold starts a single MCU active window comprising sequential SPI reading and RF transmission phases, after which the MCU returns to the system-off state.}
    \label{fig:fifo_synchronized_on_demand}
\end{figure}

The FIFO interrupt threshold was configurable from 1 to 32 words, thereby emulating different levels of transmission delay and real-time demands. A smaller FIFO threshold results in more frequent transmissions, whereas a larger threshold leads to more burst-like and discrete transmissions. To quantify its power savings, the proposed ESB on/off operation was compared with the BLE continuous-connection mode and ESB standby as baselines.

For BLE communication, the system was configured to operate in continuous-connection mode. The communication parameters were kept consistent with those used in the previous 7.5 ms experiment. In this configuration, whenever a FIFO interrupt was triggered, the SPI interface promptly retrieved the buffered data and transmitted it via the BLE link during the available connection event~\cite{BluetoothCoreSpecV62}. The BLE sleep--wake (on/off) operation was not adopted because the advertising and reconnection procedures introduce a relatively long and highly variable wake-up delay. This delay may approach the FIFO full-buffer interval of 250 ms~\cite{max30003_datasheet}, thereby increasing the risk of data overflow. 

For ESB communication, the system was configured to operate under two different modes. The first mode is the standby mode, in which ESB remains in an active state and is ready to transmit data at any time. Once a FIFO interrupt is triggered, the MCU immediately retrieves the buffered data and transmits it through the ESB link using the 4 Mbps PHY. This operating mode is comparable to the continuous-connection configuration adopted for BLE in the previous experiments.

The second operating mode is the proposed ESB on-demand mode, which implements on/off operation by placing the MCU in the system-off state between FIFO-triggered transmissions. Upon a FIFO interrupt, a falling edge on the general-purpose input/output (GPIO) pin awakens the MCU and initiates the initialization phase. The MCU then retrieves the buffered samples through an SPI batch read and transmits them using the ESB 4 Mbps PHY. After receiving an ACK from the ESB receiver, the MCU returns to the system-off state. By restricting MCU and radio activity to short FIFO-triggered communication windows, the proposed mode reduces both processing and communication idle power consumption.

\section{Protocol Comparison Results}


\subsection{Sleep--Wake Performance}
\begin{figure}[!htbp]
    \centering
    \includegraphics[width=0.76\columnwidth]{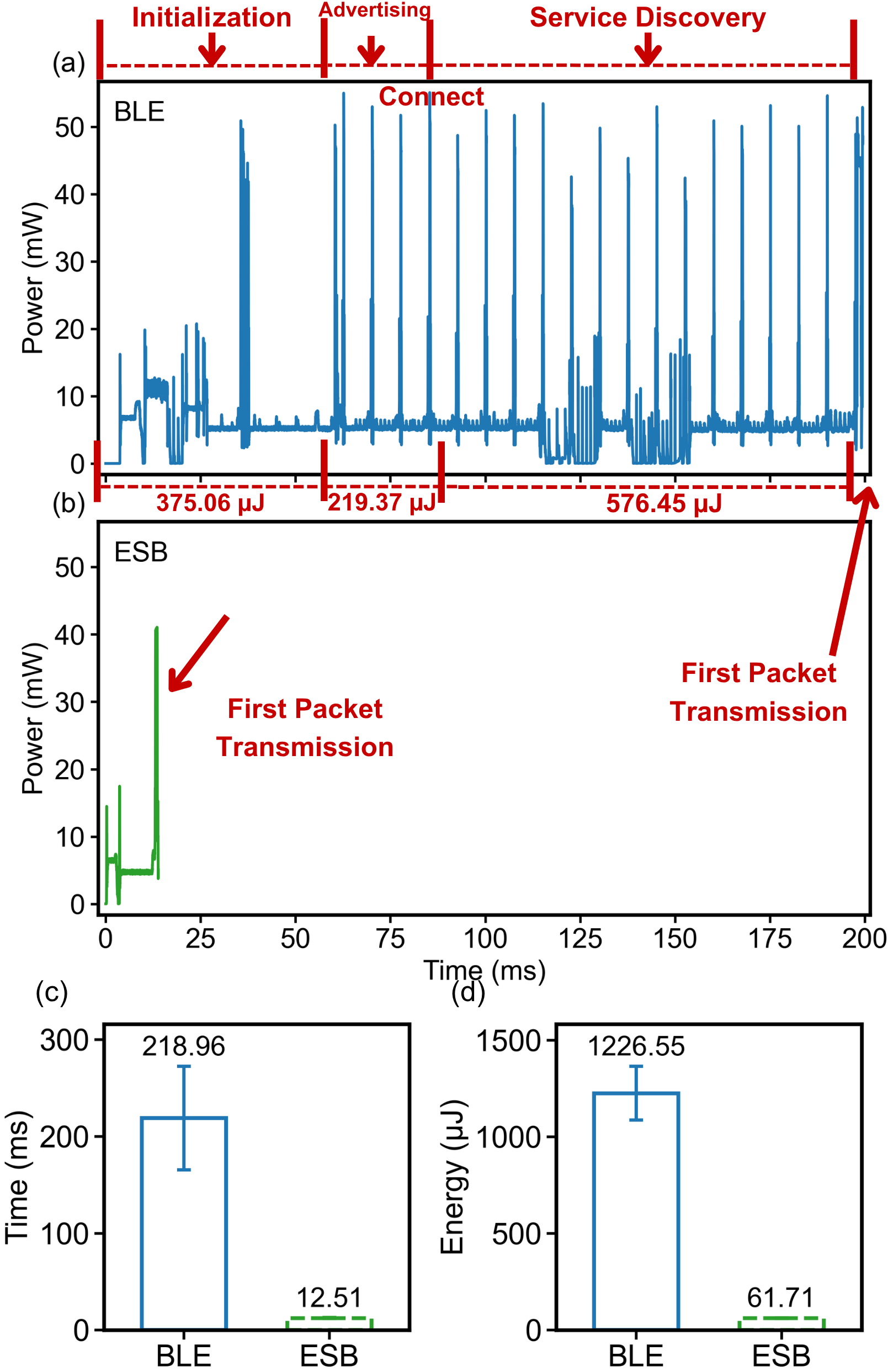}
    \caption{Wake-up and first-packet transmission of BLE and ESB: power profiles for (a) BLE and (b) ESB, and comparisons of (c) wake-up time and (d) energy consumption.}
    \label{fig:ble_esb_wakeup_first_packet}
\end{figure}

The comparison between BLE and ESB under sleep-wake communication is shown in Fig.~\ref{fig:ble_esb_wakeup_first_packet}. It illustrates the sequence in which each protocol wakes from sleep, initializes the system, transmits one data packet, and subsequently returns to sleep. 

Fig.~\ref{fig:ble_esb_wakeup_first_packet}(a) shows that, after BLE wakes up and completes the necessary system initialization, it must still undergo advertising, scan response, connection, and service discovery before data transmission can begin. The energy consumed by advertising was measured to be 219.37 $\mu$J, while service discovery accounted for an additional 576.45 $\mu$J. These results indicate that BLE incurs substantial time and energy overhead before the first packet can be transmitted. In contrast, as shown in Fig.~\ref{fig:ble_esb_wakeup_first_packet}(b), ESB behaves markedly differently. After simpler system initialization, ESB can immediately begin transmitting data packets without any additional connection procedure, thereby avoiding the large pre-transmission overhead observed in BLE.

This difference is further quantified in Fig.~\ref{fig:ble_esb_wakeup_first_packet}. As shown in Fig.~\ref{fig:ble_esb_wakeup_first_packet}(c), BLE requires an average warm-up time of 218.96 ms, whereas ESB requires only 12.51 ms, indicating a substantially shorter transition to the communication state. Correspondingly, Fig.~\ref{fig:ble_esb_wakeup_first_packet}(d) shows that BLE consumes 1226.55 $\mu$J during warm-up, while ESB consumes 61.71 $\mu$J. These results consistently demonstrate that ESB achieves approximately one-twentieth the wake-up time and wake-up energy consumption of BLE, highlighting its superior efficiency for frequent wake-up operations.

These results indicate that, for applications involving frequent wake-up and sleep cycles, ESB can shorten the response time after each wake-up event while also significantly reducing the overall energy consumption.

\subsection{Bidirectional Performance}

\begin{figure}[!htbp]
    \centering
    \includegraphics[width=0.72\columnwidth]{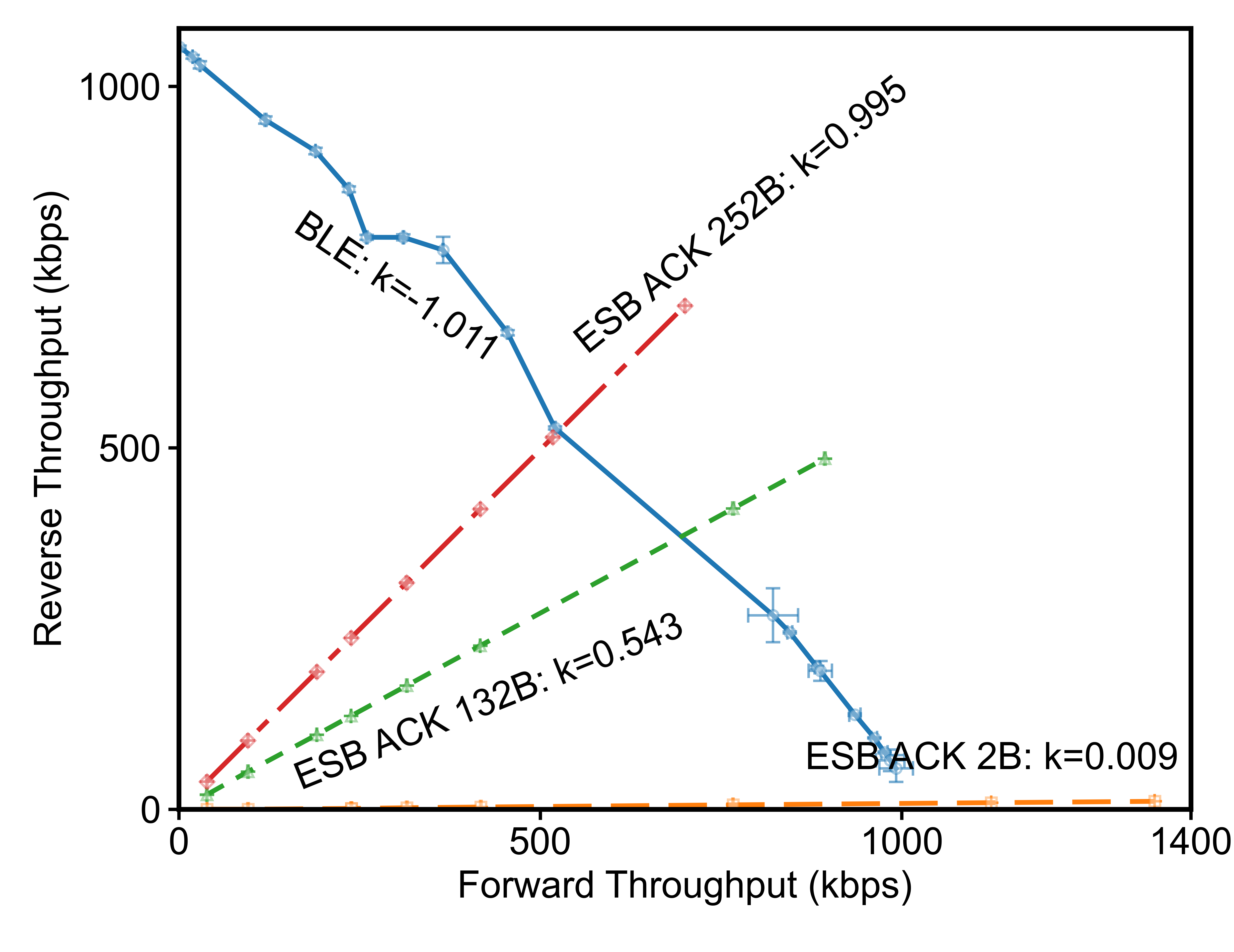}
    \caption{Bidirectional throughput trade-offs of BLE and ESB. Forward and reverse denote peripheral-to-central (transmitter-to-receiver) transmission and the opposite direction, respectively. The ESB curves correspond to ACK payloads of 2, 132, and 252 bytes; k denotes the fitted slope.}
    \label{fig:bidir}
\end{figure}

Fig.~\ref{fig:bidir} compares the bidirectional throughput of BLE and ESB under the controlled-rate methodology. For BLE, the peripheral-to-central and central-to-peripheral throughputs exhibit a strong negative correlation (k = -1.011). As the forward throughput increases from approximately 1 kbps to 992.20 kbps, the reverse throughput decreases from about 1054.10 kbps to 56.29 kbps. The operating points follow an approximately linear trade-off, while the mean aggregate throughput remains approximately 1100 kbps. This indicates that maximizing unidirectional throughput in BLE requires reducing traffic in the opposite direction.

In contrast, ESB demonstrates a fundamentally different behavior governed by ACK payload size. With 2-byte ACK packets, the reverse throughput remains negligible (k = 0.009); at the displayed forward-throughput endpoint of 1350 kbps, the reverse throughput is only 10.84 kbps, yielding an aggregate throughput of 1360.84 kbps. Increasing the ACK size to 132 bytes significantly enhances the reverse throughput (k = 0.543). At the displayed endpoint of 893.65 kbps in the forward direction, the reverse throughput reaches 484.97 kbps, corresponding to an aggregate throughput of 1378.62 kbps. When the ACK size is further increased to 252 bytes, the forward and reverse throughputs become nearly symmetric (k = 0.995), reaching 700 kbps and 696.40 kbps, respectively, at the displayed endpoint and yielding an aggregate throughput of 1396.40 kbps.

It is worth noting that reverse throughput in ESB may experience data loss, as the ACK packet itself does not support ACK or retransmission mechanisms. Under normal laboratory conditions, the packet loss rate reached approximately 22.6\% when the received signal strength indicator (RSSI) was around -75 dBm and a 252-byte ACK payload was used for data return. In addition, the ACK size in ESB does not simply redistribute a fixed throughput budget, but directly determines the payload efficiency of each transaction. Moreover, achieving maximum unidirectional throughput requires minimizing the payload size at the peer node while operating at saturation speed.

The results highlight the fundamental limitation of ESB in bidirectional communication. BLE exhibits a shared-capacity trade-off between the two directions, whereas ESB relies on forward-triggered reverse transmission. Consequently, BLE provides symmetric control under constrained shared capacity, while ESB achieves higher throughput at the cost of reverse-link dependency.

\subsection{Reliability Performance}

\begin{figure}[!htbp]
    \centering
    \includegraphics[width=0.78\columnwidth]{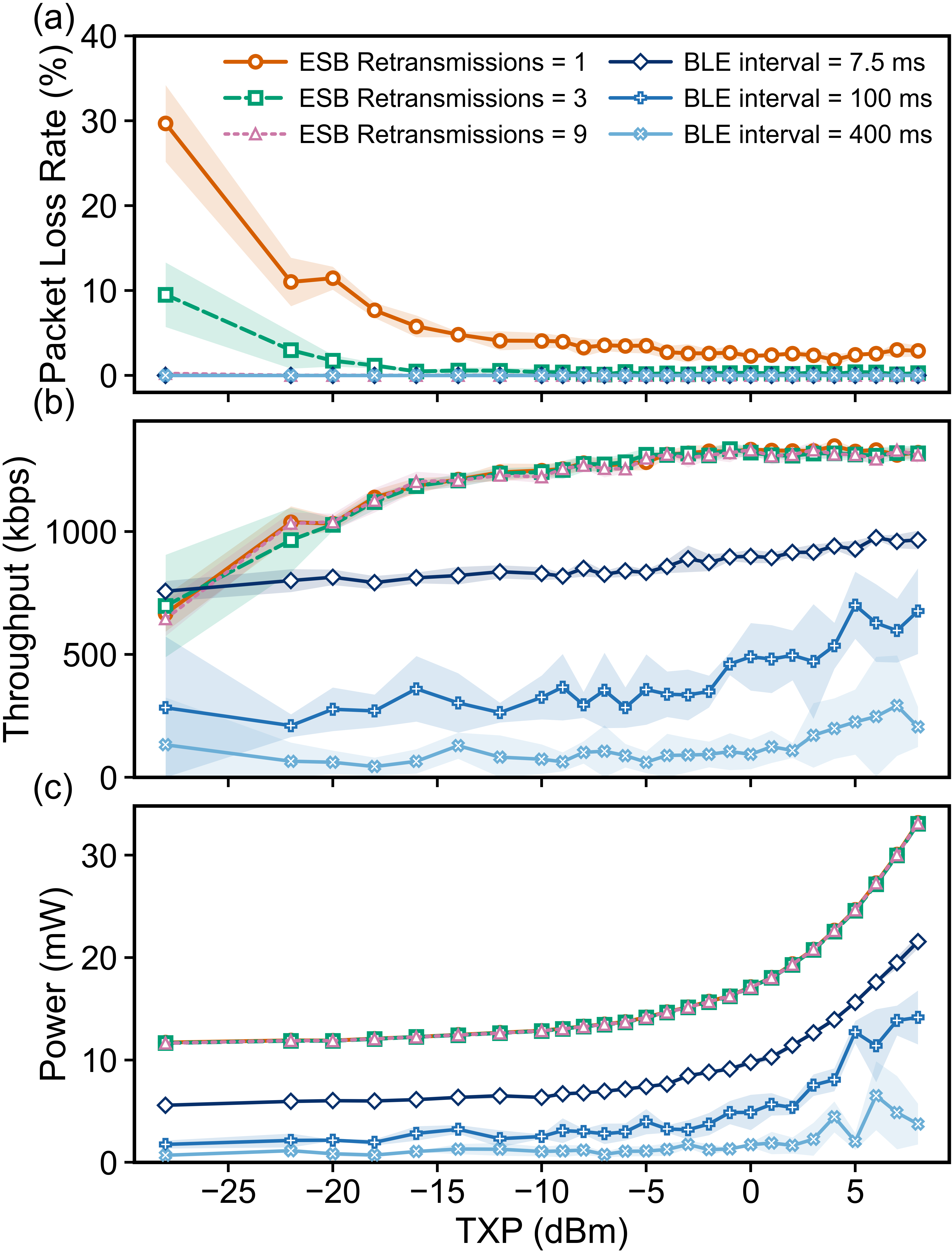}
\caption{Communication performance versus TXP for ESB with 1, 3, and 9 retransmissions and BLE with connection intervals of 7.5, 100, and 400 ms. (a) Packet loss rate, (b) throughput, and (c) power consumption.}
    \label{fig:txp_protocol_comparison}
\end{figure}

As shown in Fig.~\ref{fig:txp_protocol_comparison}(a), BLE maintained a measured packet loss rate of 0\% across all tested TXP settings for connection intervals of 7.5, 100, and 400 ms. This result indicates that the BLE Link Layer acknowledgment and retransmission procedure provides strong robustness under the investigated interference conditions. In contrast, the reliability of ESB was governed by both the effective link margin and the configured retransmission count. With one retransmission, the packet loss rate reached 29.68\% at the minimum TXP of -28 dBm and decreased to 0\% at 8 dBm, demonstrating that a stronger signal substantially improves ESB reliability in the presence of interference. The retransmission count provided a second mechanism for improving ESB reliability. At -28 dBm, increasing the retransmission count from 1 to 3 reduced the packet loss rate from 29.68\% to 9.50\%, while 9 retransmissions further reduced it to 0\%. Thus, additional retransmission opportunities can compensate for a weak link and reduce the ESB packet loss rate to zero.

However, the reduction in packet loss was not achieved without cost for ESB. As shown in Fig.~\ref{fig:txp_protocol_comparison}(b), additional retransmission attempts occupy radio time and increase the queueing delay of subsequent packets, thereby reducing the effective throughput when the retransmission overhead becomes dominant. At -28 dBm, the throughput decreased sequentially from 697.0 kbps with one retransmission to 666.0 kbps with three retransmissions and 644.6 kbps with nine retransmissions. Meanwhile, the corresponding packet loss rate decreased from 29.68\% to 9.50\% and finally to 0\%. These results reveal a fundamental reliability--throughput trade-off: a small retransmission count may preserve transmission opportunities at the expense of data loss, whereas additional retransmission improves delivery reliability but increases airtime occupancy and packet queueing.

For BLE, Fig.~\ref{fig:txp_protocol_comparison}(b) shows a clear increase in throughput as the connection interval decreases. This behavior arises because a shorter connection interval schedules connection events more frequently, reducing the waiting time between transmission opportunities and thereby increasing the number of packets delivered per unit time. At -28 dBm, reducing the connection interval from 400 to 100 and 7.5 ms increased the throughput from 132.33 to 283.0 and 756.8 kbps, respectively. At 8 dBm, the corresponding throughputs increased from 205.2 to 676.0 and 965.4 kbps. Thus, a short BLE connection interval reduces scheduling latency and increases the available transmission opportunities, thereby improving effective throughput, while the packet loss rate remains at 0\% under the tested interference conditions, indicating strong robustness to interference.

Fig.~\ref{fig:txp_protocol_comparison}(c) further shows that the three ESB retransmission configurations consumed nearly identical power across the tested TXP range. Under saturated traffic, the ESB radio remained continuously active: whether a transmission opportunity carried a new packet or a retransmission, a similar amount of energy was required to transmit the packet. Consequently, changing the retransmission limit primarily affected packet delivery and queueing rather than the average radio duty cycle. For BLE, the average power increased as the connection interval decreased because more frequent connection events activated the radio more often and shortened the low-power idle periods between events. ESB nevertheless consumed more power than BLE in this continuous-transmission experiment because it lacks an equivalent connection-interval-based duty-cycling mechanism and therefore maintained a substantially higher radio activity.

\section{PHY Comparison Results}

Following the protocol-level comparison, we next evaluate the ESB 2 Mbps and 4 Mbps PHYs to identify the configuration best suited to on-demand sensing.

\begin{figure}[!htbp]
    \centering
    \includegraphics[width=0.76\columnwidth]{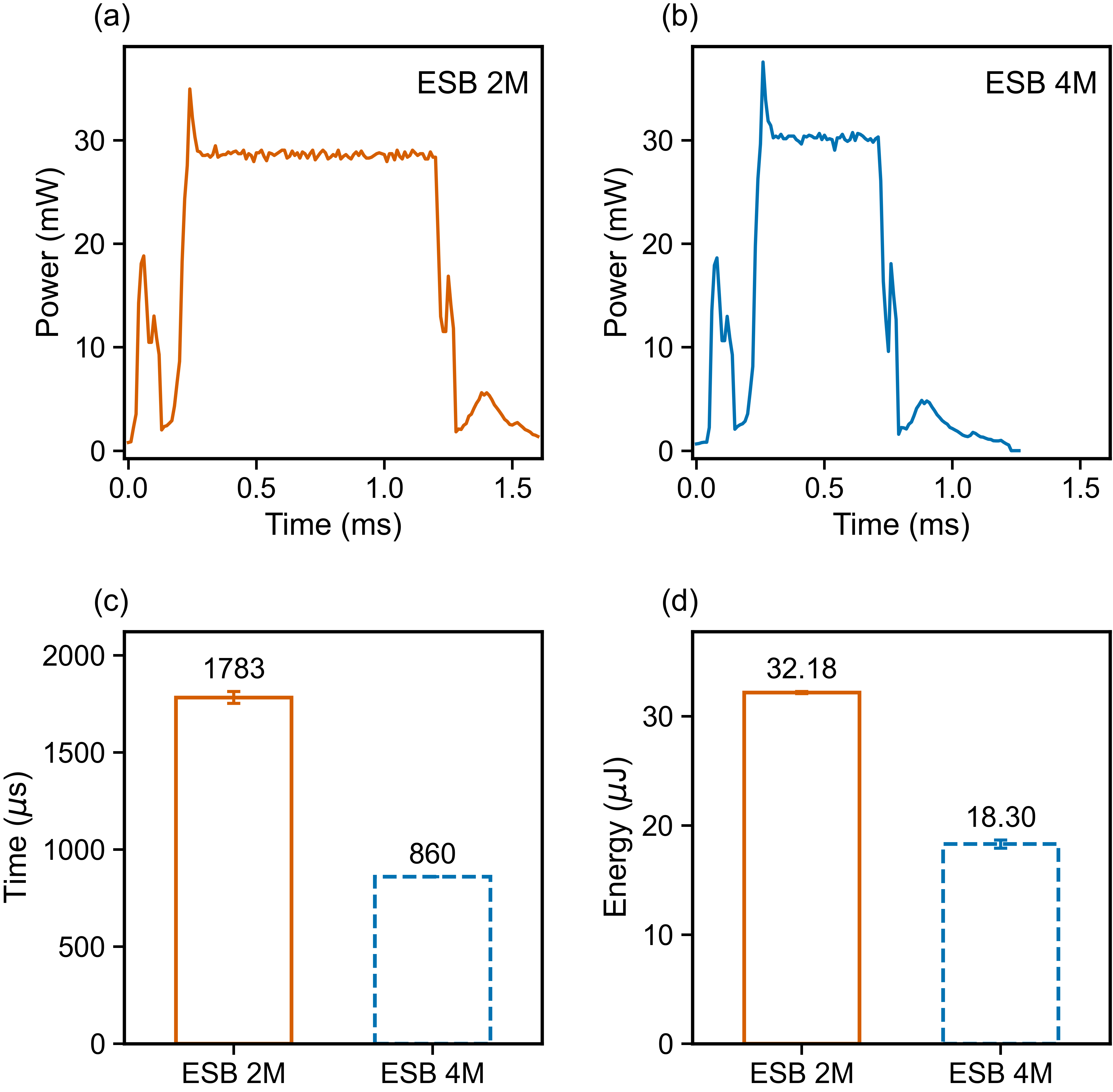}
    \caption{Single-packet performance of the ESB 2 Mbps and 4 Mbps PHYs: power profiles for (a) 2 Mbps and (b) 4 Mbps, and comparisons of (c) transmission time and (d) energy consumption.}
    \label{fig:esb_phy_single_packet_combined}
\end{figure}

Fig.~\ref{fig:esb_phy_single_packet_combined}(a) and (b) show the power profiles for transmitting one packet using the ESB 2 Mbps and 4 Mbps PHYs, respectively. Both profiles exhibit a similar transmission sequence, including radio startup, a relatively stable active-transmission stage, and radio shutdown. The principal difference is the duration of the active stage: the higher PHY rate shortens the packet airtime while maintaining a comparable instantaneous power level. This difference is quantified in Fig.~\ref{fig:esb_phy_single_packet_combined}(c). The mean single-packet transmission time decreased from approximately 1783~$\mu$s with the 2 Mbps PHY to 860~$\mu$s with the 4 Mbps PHY, corresponding to a reduction of approximately 52\%. The shorter transmission duration also reduced the energy consumption from 32.18 to 18.30~$\mu$J per packet, as shown in Fig.~\ref{fig:esb_phy_single_packet_combined}(d), representing an energy reduction of approximately 43\%. These results demonstrate that the ESB 4 Mbps PHY transmits the same packet in approximately half the time and with substantially lower energy than the 2 Mbps PHY, making it more suitable for short, event-driven transmissions in on-demand sensing systems.

\begin{figure}[!htbp]
    \centering
    \includegraphics[width=0.78\columnwidth]{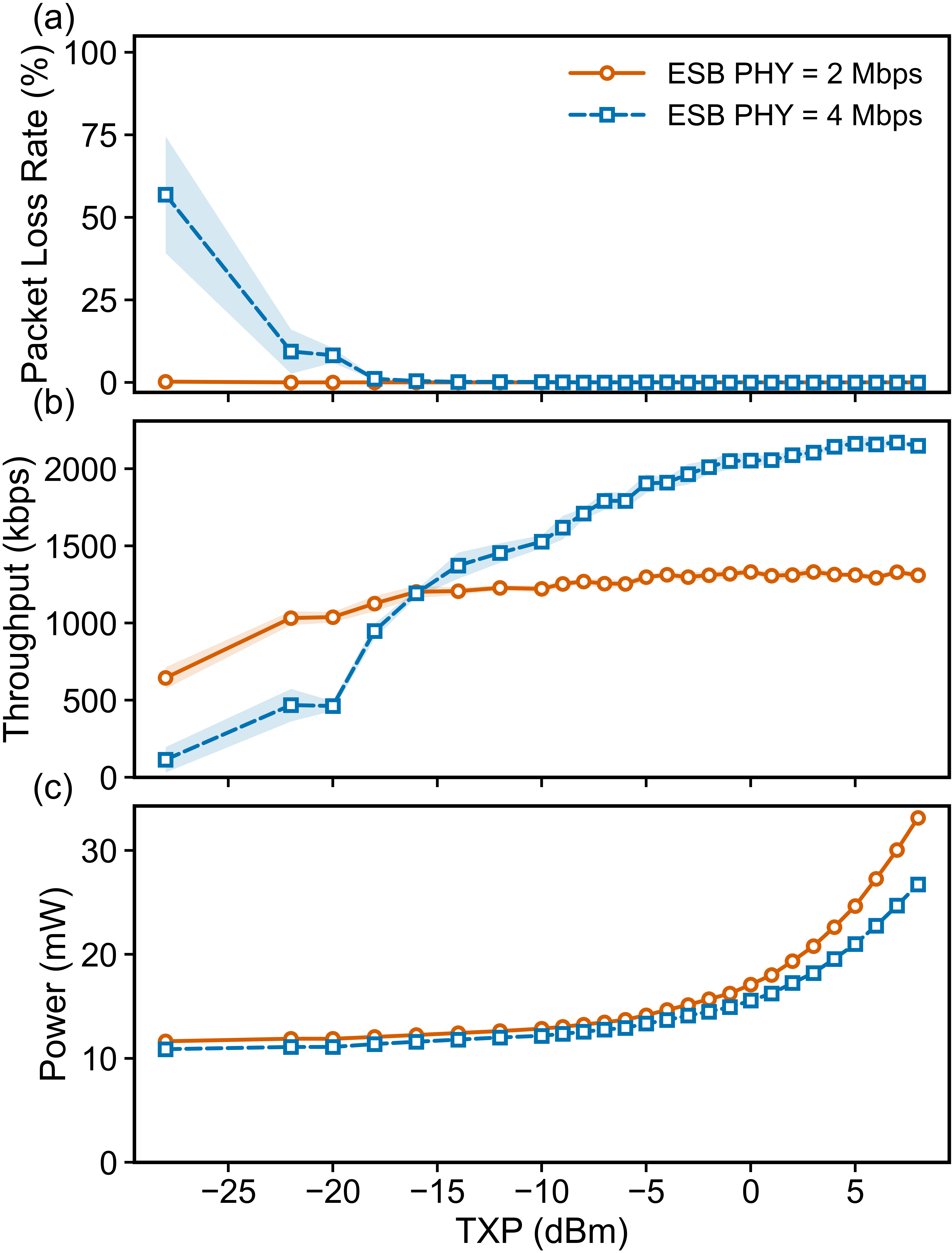}
\caption{Effect of TXP on ESB communication using the 2 Mbps and 4 Mbps PHYs. (a) Packet loss rate, (b) throughput, and (c) power consumption. }
    \label{fig:esb_phy_2m_4m_txp_comparison}
\end{figure}

Fig.~\ref{fig:esb_phy_2m_4m_txp_comparison}(a) and (b) compare the two PHYs under continuous packet transmission. At high TXP, the 4 Mbps PHY achieved a peak throughput of approximately 2.15 Mbps, compared with approximately 1.31 Mbps for the 2 Mbps PHY. This throughput advantage diminished under weak-link conditions because the higher-rate PHY had a smaller effective link margin. At -28 dBm, the packet loss rate of the 4 Mbps PHY exceeded 50\% even with nine retransmissions, whereas the packet loss rate of the 2 Mbps PHY remained at 0\%. Therefore, the higher throughput of the 4 Mbps PHY is accompanied by reduced robustness at low TXP. 

As shown in Fig.~\ref{fig:esb_phy_2m_4m_txp_comparison}(c), the 4 Mbps PHY also consumed slightly less average power than the 2 Mbps PHY across the tested TXP range. At 8 dBm, for example, the measured power decreased from approximately 33 to 27 mW. This reduction can be attributed to the shorter packet airtime of the higher-rate PHY. During continuous transmission, the 4 Mbps PHY completes each over-the-air transmission more rapidly and therefore spends a larger fraction of time in the lower-power inter-packet turnaround required to prepare and switch to the next packet, rather than in the active transmit state. Consequently, the 4 Mbps PHY improves both throughput and energy efficiency when the link margin is sufficient, while the 2 Mbps PHY remains preferable when communication robustness under weak-signal conditions is the primary requirement.

On-demand sensing applications typically operate over short communication distances with a relatively strong link margin and predominantly involve unidirectional data transmission, as in ILR systems. Therefore, the ESB 4 Mbps PHY with nine retransmissions was selected for the subsequent on-demand sensing evaluation.

The optimized ESB configuration is subsequently incorporated into a practical sensing system to investigate whether the protocol-level advantages translate into system-level energy savings.

\section{System Evaluation Results}

\begin{figure*}[!htbp]
    \centering
    \includegraphics[width=0.99\textwidth]{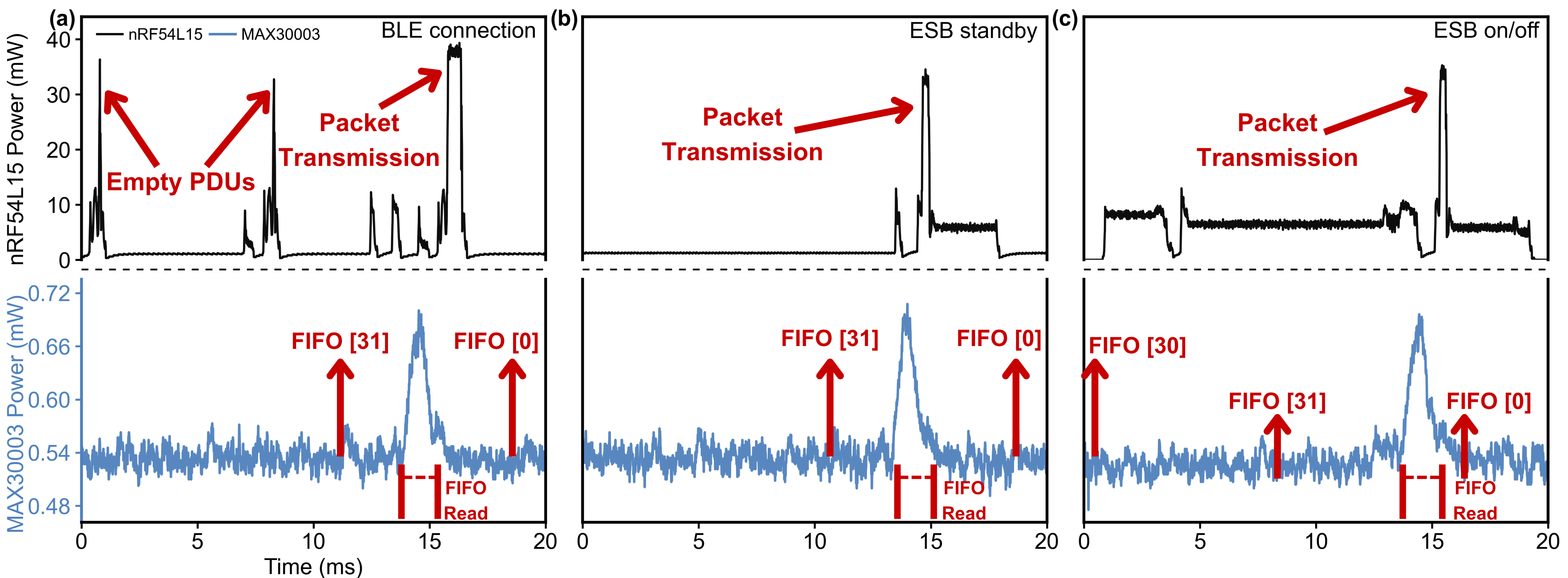}
    \caption{Power consumption of the nRF54L15 and MAX30003 in different wireless modes: (a) BLE connection, (b) ESB standby, and (c) ESB on/off.}
    \label{fig:fifo_power_profile}
\end{figure*}

Fig.~\ref{fig:fifo_power_profile} illustrates the continuous data streaming behavior of the ILR during abnormal ECG monitoring. Subfigures (a)--(c) present the instantaneous power consumption of both the MCU nRF54L15 and the sensor MAX30003 when the maximum FIFO interrupt threshold is selected. For both BLE and ESB standby modes, a FIFO threshold of 32 words was configured. In contrast, a threshold of 31 words was adopted for the ESB on/off mode to prevent potential data overflow caused by the relatively long warm-up and initialization duration.

Fig.~\ref{fig:fifo_power_profile}(a) illustrates the instantaneous power profile under the BLE connection mode. It can be observed that empty Protocol Data Units (PDUs) are periodically transmitted in each connection interval to maintain the link. When the FIFO index reaches 31, indicating that 32 samples have been accumulated, a FIFO interrupt is triggered. Immediately after sensor data are acquired via SPI (FIFO read), data transmission is initiated through the BLE link. Upon completion of the transmission, the FIFO index resets. This timing sequence prevents FIFO overflow and data collision, thereby ensuring data integrity during continuous streaming.

Fig.~\ref{fig:fifo_power_profile}(b) for ESB standby exhibits a trend similar to that observed under the BLE connection mode. However, it is notable that the packet duration is significantly shorter because ESB uses the 4 Mbps PHY, whereas BLE uses the 2 Mbps PHY. Another distinction is that, after data transmission, the ESB system requires additional time before entering the standby state. This interval corresponds to the completion of the ESB transaction, during which the radio switches from transmit (TX) to receive (RX) mode to wait for the ACK, performs any required retransmissions, and then returns to the idle state.

Fig.~\ref{fig:fifo_power_profile}(c) presents the transient power profile when employing the ESB sleep--wake (on/off) operation. It can be observed that the MCU is awakened once 31 samples (FIFO[30]) have accumulated in the FIFO buffer. After waking up, the MCU performs system initialization before continuing data acquisition. Subsequently, after completing the FIFO read operation, the ESB link transmits the buffered data in a burst manner, and the system then returns to the sleep state. 

Compared with the shorter transmission interval observed in Fig.~\ref{fig:fifo_power_profile}(b), the longer inactive duration in on/off mode is mainly attributed to the additional processing overhead associated with transitioning into the system off state. This includes register configuration, waiting for the radio state machine to return to the DISABLED state, and clearing the Distributed Programmable Peripheral Interconnect (DPPI) channels and timer resources.

From a long-term system-level power consumption perspective, adjusting the FIFO wake-up threshold from 1 to 32 results in only marginal variations in the MAX30003 power profile across the three communication schemes, with the average power consumption converging to approximately 0.5 mW. This behavior is primarily attributed to the fixed sensor sampling rate of 128 Hz. In addition, the energy required for FIFO readout is strongly correlated with the number of samples retrieved. Consequently, modifying the wake-up interval does not introduce a significant change in the sensor's average power consumption.

\begin{figure}[!htbp]
    \centering
    \includegraphics[width=0.72\columnwidth]{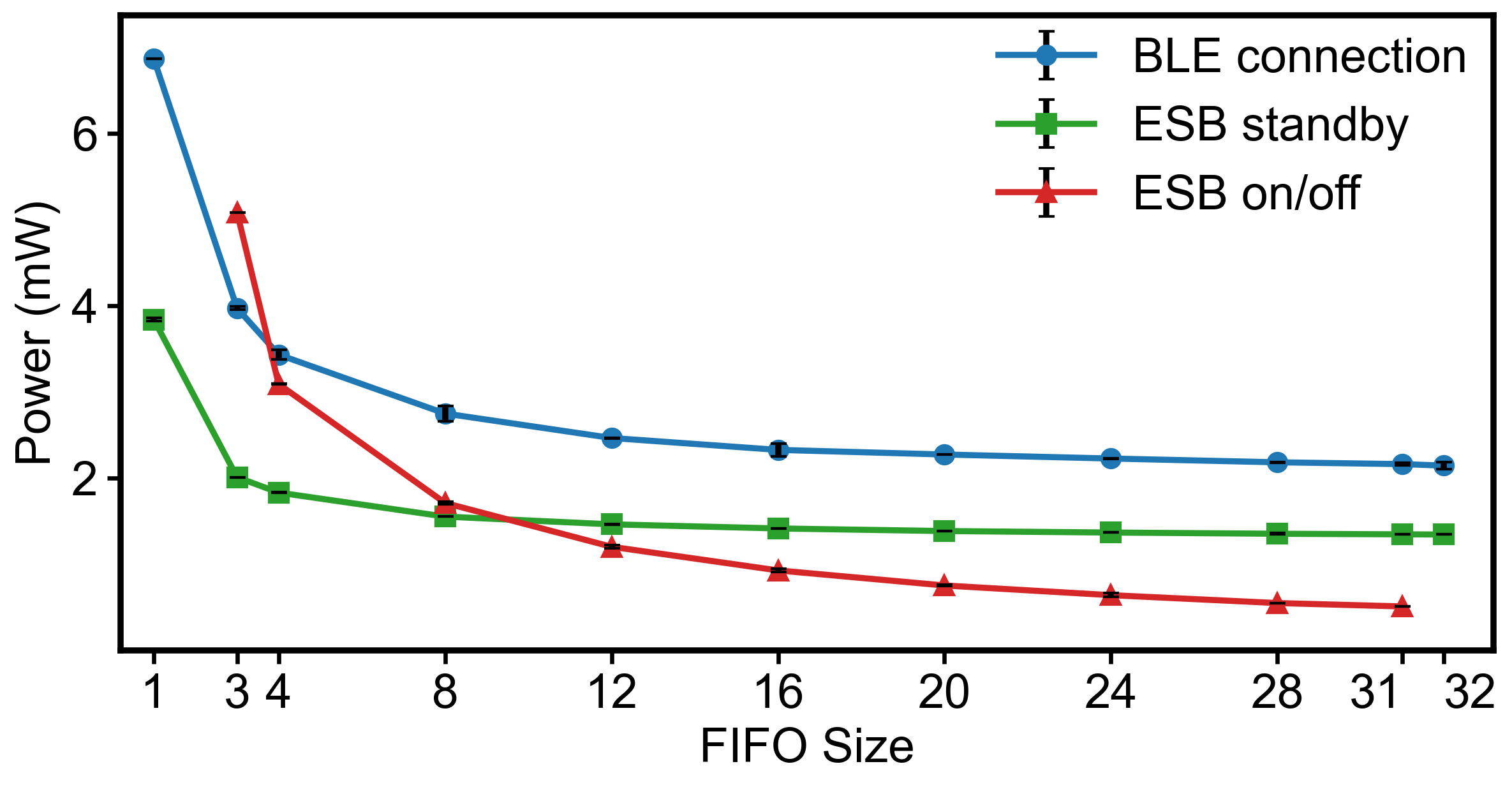}
    \caption{Power consumption of the nRF54L15 as a function of FIFO size under BLE connection, ESB standby, and ESB on/off communication modes.}
    \label{fig:nrf54l15_fifo_power}
\end{figure}

Because the sensor power consumption is low and exhibits minimal variation, the MCU-side power optimization becomes the primary focus. Fig.~\ref{fig:nrf54l15_fifo_power} illustrates the variation in MCU power consumption as a function of the FIFO interrupt threshold. For all three communication schemes, the power consumption decreases as the FIFO threshold increases. This trend arises because the FIFO buffer enables multiple samples to be aggregated and transmitted in a single communication event, thereby reducing the overhead associated with frequent radio activation. 

However, the three methods exhibit distinct trends. Among them, the BLE connection mode shows the highest overall power consumption. When the FIFO interrupt threshold is set to 1, the MCU power consumption reaches approximately 6.6 mW, and even at the maximum FIFO threshold, the power consumption remains around 2.1 mW. This behavior is attributed to the need for BLE to continuously exchange PDUs to maintain the connection. Furthermore, BLE operates with a 2 Mbps PHY in this configuration, leading to relatively higher transmission energy per bit.

For the ESB standby mode, the overall power consumption is consistently lower than that of BLE across all FIFO interrupt thresholds. Specifically, the MCU power consumption decreases from approximately 3.8 mW at FIFO = 1 to about 1.3 mW at FIFO = 32, demonstrating a substantial improvement compared with BLE. This enhancement mainly arises because ESB operates as a lightweight, connectionless protocol that does not require additional energy expenditure to maintain link connectivity. Moreover, the higher PHY data rate further reduces the transmission energy per bit, contributing to improved overall energy efficiency.

\begin{figure}[!htbp]
    \centering
    \includegraphics[width=0.72\columnwidth]{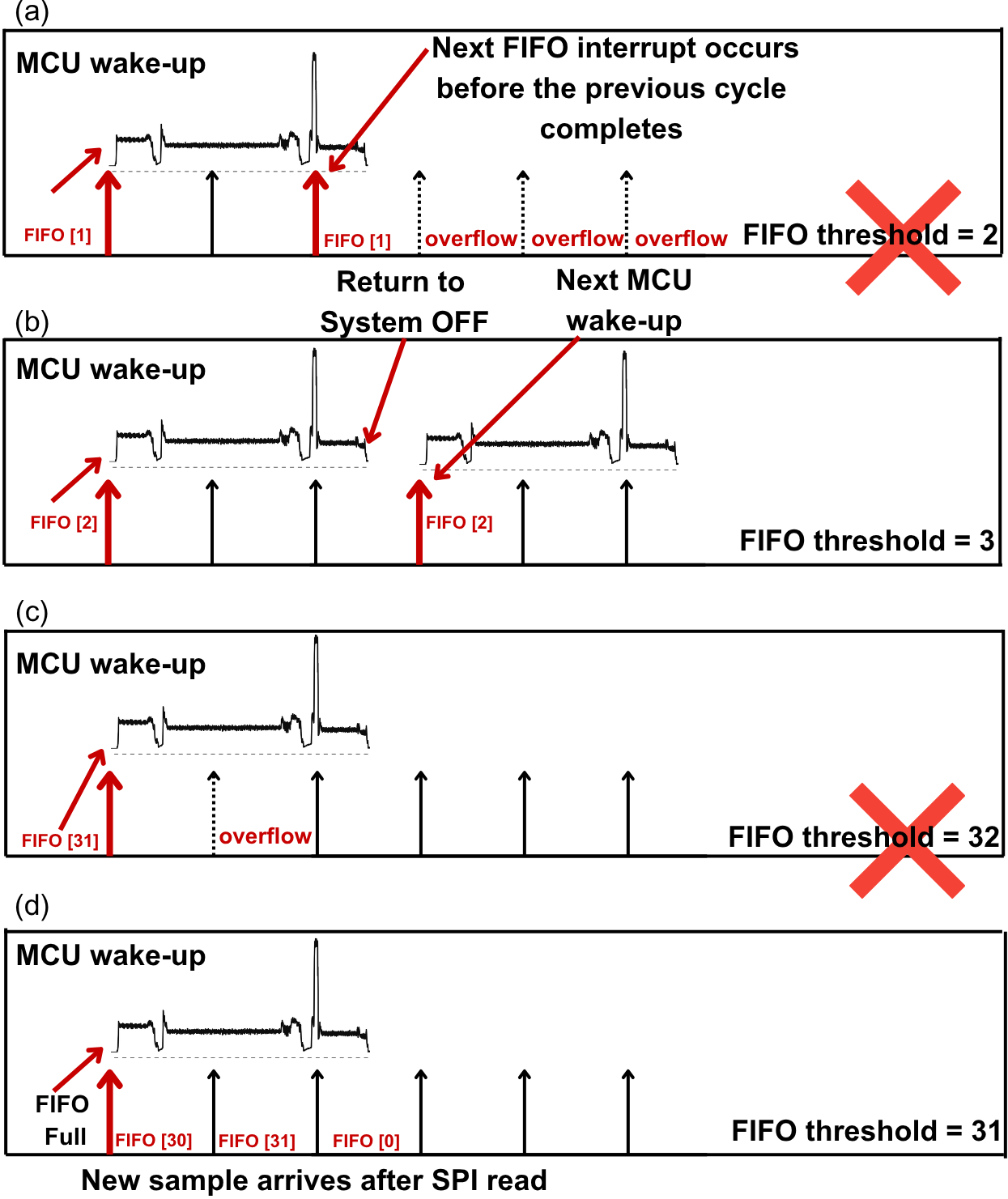}
    \caption{Timing Constraints Governing Reliable FIFO-Triggered Sleep--Wake Operation at an ECG sampling rate of 128 Hz.}
    \label{fig:esb_onoff_fifo_threshold}
\end{figure}

For the ESB on/off mode, the feasible FIFO interrupt threshold is limited to 3--31, as illustrated in Fig.~\ref{fig:esb_onoff_fifo_threshold}. The MAX30003 sampling frequency was 128 Hz, corresponding to one ECG sample every $1/128$ s, or 7.8125 ms. With a threshold of 2, consecutive wake-up events are separated by only 15.625 ms, as shown in Fig.~\ref{fig:esb_onoff_fifo_threshold}(a). This interval is shorter than the time required to complete MCU wake-up, initialization, SPI readout, ESB transmission and acknowledgment, and the return to the system-off state. Consequently, the next FIFO interrupt can occur before the previous cycle has completed, causing overlapping wake-up cycles and persistent FIFO overflow. Increasing the threshold to 3 extends the interval to 23.4375 ms, allowing the MCU to complete the sequence and return to the system-off state before the next interrupt, as shown in Fig.~\ref{fig:esb_onoff_fifo_threshold}(b).

The upper limit is imposed by the 32-sample FIFO capacity. A threshold of 32 triggers the interrupt only after the FIFO is full and therefore leaves no free entry. Because another sample arrives 7.8125 ms later, it may be generated during MCU wake-up and initialization before the SPI read releases FIFO space, resulting in overflow, as shown in Fig.~\ref{fig:esb_onoff_fifo_threshold}(c). In contrast, a threshold of 31 leaves one empty entry for the sample arriving during initialization; the subsequent SPI read then clears the buffered samples, as shown in Fig.~\ref{fig:esb_onoff_fifo_threshold}(d). Therefore, thresholds from 3 to 31 provide the timing margin required for reliable ESB on/off operation.

At smaller FIFO thresholds, the power consumption of the ESB in the on/off mode is higher than that of the ESB in the standby mode. This is mainly because frequent wake-up events introduce significant initialization overhead, which dominates the overall energy consumption. However, as the FIFO threshold increases and the initialization frequency decreases, the advantage of the on/off strategy becomes progressively more evident. In particular, when the FIFO threshold reaches its maximum value of 31, the MCU power consumption is reduced to approximately 0.5 mW, representing a substantial improvement compared with both the ESB standby mode and the BLE connection mode.

This observation indicates that, for applications involving transmission of abnormal ECG waveforms, adopting the ESB on/off communication strategy can reduce the total system power consumption by approximately 60\% compared with the BLE continuous-connection mode. Consequently, this approach has the potential to extend the operational battery lifetime of ILR systems in this application.

\section{Discussion}
The protocol comparison shows that communication architecture should be selected according to the traffic pattern of the sensing application. BLE is well suited to systems that require persistent bidirectional access, standardized security, and direct interoperability with smartphones or commercial gateways. These functions justify its connection-management overhead in continuous monitoring and interactive devices. In on-demand sensing, however, long idle periods are interrupted by only brief transmission events. Establishing and maintaining a connection then extends radio activity without increasing the amount of sensing data delivered. ESB avoids this overhead and permits immediate transmission after wake-up, but its reverse link depends on a forward packet carrying an ACK payload, and it does not provide the standardized ecosystem of BLE. Neither architecture is therefore preferable for every sensing application.

Because BLE and ESB were implemented on the same Nordic platform, differences in radio hardware, clocks, antennas and power management were excluded from the comparison. The measured differences can therefore be attributed mainly to protocol operation. This hardware compatibility also permits a Nordic-based sensing node to switch between BLE and ESB through firmware changes without redesigning its RF hardware, reducing the practical barrier to using connectionless transmission when continuous connectivity is unnecessary.

The ILR experiment explains how the communication method interacts with the sensing schedule. The MAX30003 first stores multiple ECG samples in its FIFO while the MCU remains inactive. A FIFO interrupt then initiates MCU startup, SPI readout, and ESB transmission within one active period, after which the node returns to the system-off state. Transmitting several samples per wake-up distributes the fixed costs of initialization and radio activation across the buffered data. The observed system-level power reduction therefore results from both low-overhead ESB transmission and FIFO-controlled wake-up.

These findings lead to practical configuration rules. Applications requiring persistent bidirectional sessions are better supported by BLE, whereas connectionless operation is better suited to systems that buffer sensing data locally and transmit it intermittently. Within ESB, the 4 Mbps PHY reduces airtime and transmission energy when link margin is sufficient, while the 2 Mbps PHY provides greater robustness under weak-link conditions. Retransmission count should be selected according to the required delivery reliability because additional attempts reduce packet loss but consume transmission opportunities and lower effective throughput. The FIFO threshold should likewise be selected to reduce the average wake-up overhead while preserving sufficient timing and buffer margins.  Although ESB was used to implement connectionless communication in this study, the identified design principles may also inform the use of other connectionless protocols for event-driven sensing.

\section{Conclusion}

This work compared connection-oriented BLE and connectionless ESB for ultra-low-power on-demand sensing on a common Nordic hardware platform. ESB reduced wake-up latency and communication energy by avoiding connection establishment and link maintenance, while its 4 Mbps PHY further reduced packet airtime and transmission energy when channel conditions supported the higher rate. In the FIFO-triggered ILR prototype, coordinating buffered sensing with ESB sleep--wake transmission reduced total system power consumption by approximately 60\% relative to BLE continuous-connection operation. These results provide experimental guidance for selecting the communication architecture, PHY, retransmission setting, and buffering strategy according to the traffic pattern, channel condition, reliability requirement, and energy budget of the sensing application. Future work will incorporate application-layer security, improve reverse-link communication flexibility, and explore hybrid integration with standardized protocols such as BLE.

\clearpage
\bibliographystyle{IEEEtran}
\bibliography{main}

\ifincludeauthorbiographies
\begin{IEEEbiography}[{\includegraphics[width=1in,height=1.25in, clip,keepaspectratio]{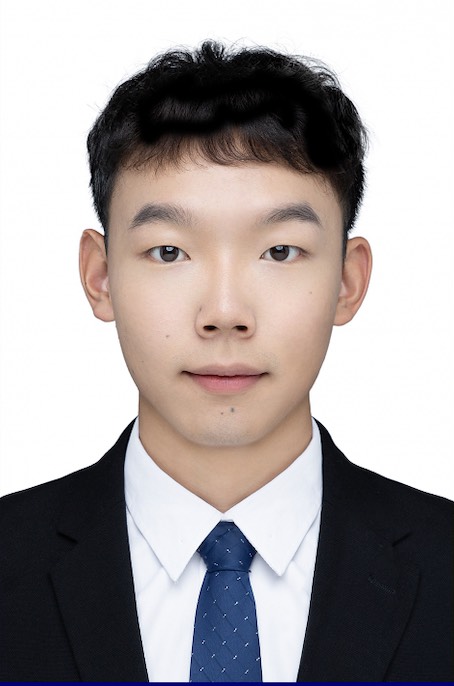}}]{Ziyao Zhou} received the B.Eng. degree in Communication Engineering from the University of Electronic Science and Technology of China, Chengdu, China, in 2023, and the M.Sc. degree in Communication Engineering from Nanyang Technological University, Singapore, in 2024. He is currently working toward a Ph.D. in the Digital, AI, Robotics, and Electronics (DARE) Lab for Translational Medicine, School of Electrical and Electronic Engineering, Nanyang Technological University. His research interests include wireless sensor and actuator networks for gastrointestinal tract applications.
\end{IEEEbiography}

\begin{IEEEbiography}[ {\includegraphics[width=1in,height=1.25in,clip,keepaspectratio]
{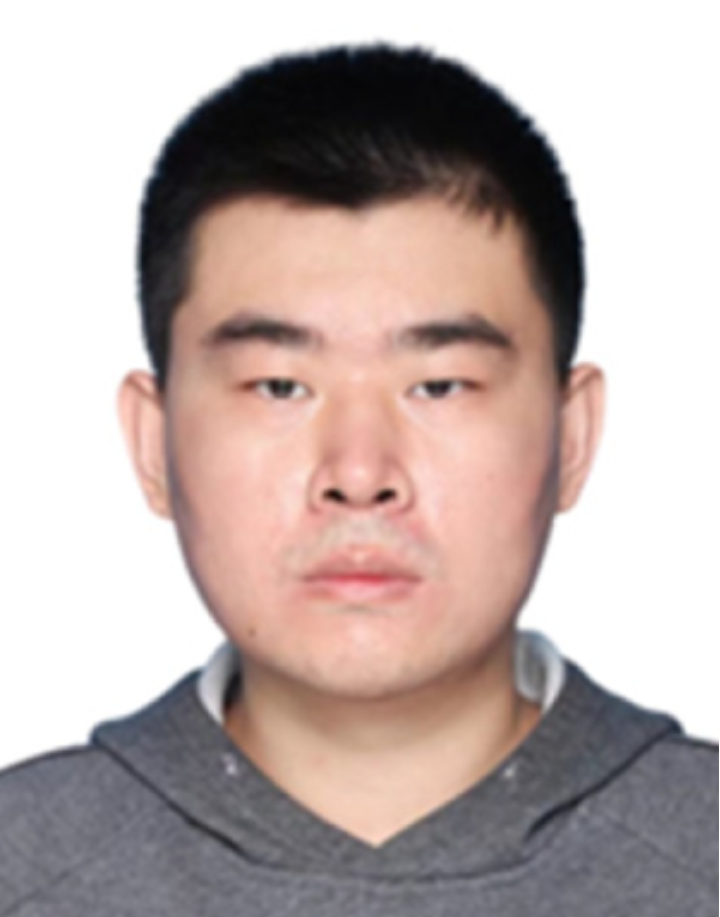}}
]{Chen Shen} received the B.Eng. degree in electrical and electronic engineering from the University of Nottingham, Ningbo, China, in 2014, and the M.Sc. and
Ph.D. degrees in electrical and electronic engineering from Nanyang Technological University, Singapore, in 2019 and 2025, respectively.
He is currently a Research Fellow in the Digital, AI, Robotics, and Electronics (DARE) Lab for Translational Medicine, School of Electrical and Electronic Engineering, Nanyang Technological University.  His research interests
include low-power ASIC design, neuromorphic computing, edge AI, and ingestible electronics.
\end{IEEEbiography}
\begin{IEEEbiography}[{\includegraphics[width=1in,height=1.25in,clip,keepaspectratio]{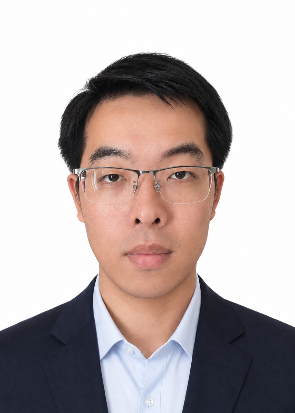}}]{Tiancheng Cao} received the B.Eng. degree with Highest Distinction and the Ph.D. degree from Nanyang Technological University (NTU), Singapore, in 2021 and 2024, respectively. He was a recipient of the Nanyang President's Graduate Scholarship, the NTU--Thomason Gold Medal, and the Best Ph.D. Thesis Award. During his doctoral studies, he also served as an attached researcher at the Institute of Microelectronics (IME), A*STAR, Singapore. He is currently a Schmidt AI in Science Fellow (Schmidt Sciences) at NTU, a Postdoctoral Fellow at Brigham and Women's Hospital/Harvard Medical School, and a Visiting Scientist at MIT. His work focuses on real-time edge-AI systems for biomedical applications, including neuromorphic and edge computing, the Internet of Medical Things (IoMT), and translational medicine.
\end{IEEEbiography}
\begin{IEEEbiography}[{\includegraphics[width=1in,height=1.25in, clip,keepaspectratio]{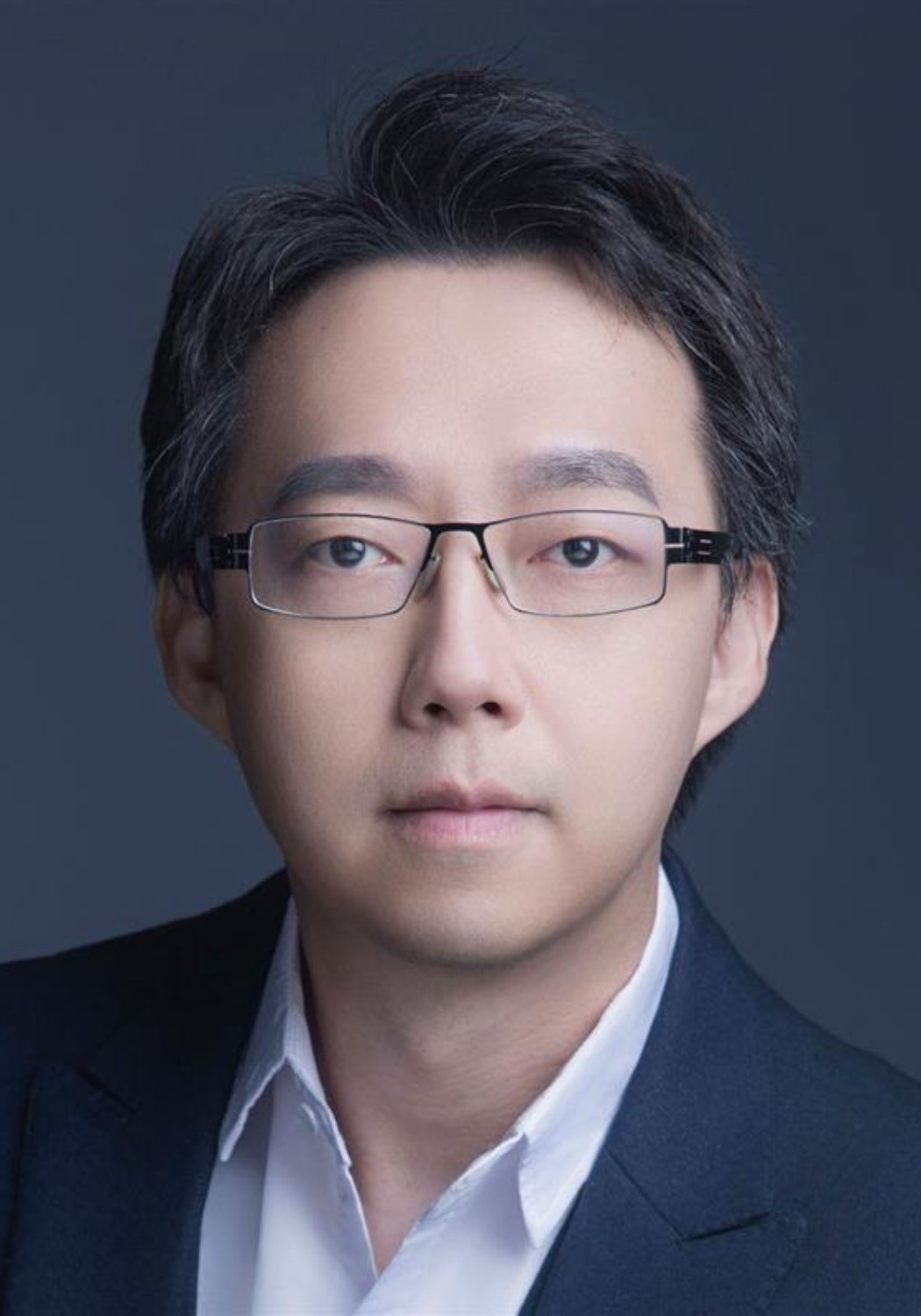}}]{Hen-Wei Huang} received the B.S. and M.S. degrees in Mechanical Engineering from National Taiwan University, Taiwan, in 2011 and 2012, respectively, and the Ph.D. degree in Robotics from ETH Zürich, Switzerland, in 2018.
He was a faculty member in Medicine at Harvard Medical School and an Associate Scientist at Brigham and Women's Hospital, Boston, USA, from 2021 to 2024, where he contributed to the development of advanced biomedical sensing and robotic systems.
He previously worked as an R\&D Engineer at TSMC (2012--2013) and MedSense Inc. (2013--2014), focusing on advanced sensing and electronic packaging technologies.
He is currently an Assistant Professor at the School of Electrical and Electronic Engineering, with a joint appointment at the Lee Kong Chian School of Medicine, Nanyang Technological University (NTU), Singapore, since 2024. He is also the Director of the DARE (Digital, AI, Robotics, and Electronics) Laboratory for Translational Medicine.
His research focuses on in vivo wireless sensor networks, ingestible and implantable electronics, robotic-assisted drug delivery, and translational medical technologies, with the goal of enabling next-generation minimally invasive diagnostics and therapeutics.
\end{IEEEbiography}
\fi



\end{document}